\documentclass[11pt]{amsart}
\usepackage{amssymb,amsmath}
\usepackage{latexsym}
\usepackage{graphicx}
\usepackage{geometry}                % See geometry.pdf to learn the layout options. There are lots.
\usepackage{color}
\usepackage{xcolor}
\usepackage{pdfsync}
\usepackage{fancyhdr}
\usepackage[applemac]{inputenc}
\usepackage[breaklinks,colorlinks,backref]{hyperref}
\hypersetup{
    colorlinks, %set true if you want colored links
    linktoc=all, %set to all if you want both sections and subsections linked
    linkcolor=red,  %choose some color if you want links to stand out
  citecolor=hyptxt,
  urlcolor=blue}
  \hypersetup{
  citebordercolor=,
  filebordercolor=green,
  linkbordercolor=blue
}
\definecolor{hyptxt}{rgb}{0.7, 0.4, 0.9}
\usepackage{bibtopic}

\newcommand{\be}{\begin{equation}}
\newcommand{\en}{\end{equation}}

\newcommand{\bea}{\begin{eqnarray}}
\newcommand{\ena}{\end{eqnarray}}

\newcommand{\ket}[1]{|\kern.3ex#1\kern.3ex\rangle}
\newcommand{\bra}[1]{\langle\kern.3ex #1 \kern.3ex|}
\newcommand{\scalar}[2]{\langle\kern.3ex #1 \kern.3ex|\kern.3ex#2\kern.3ex\rangle}

\newcommand{\C}{\mathbb C}

\newcommand{\R}{\mathbb R}

\newcommand{\SN}{{\mathbb S}}

\newcommand{\ud}{\,\mathrm{d}}

\def\lg{\langle }
\def\rg{\rangle }

\def\vs{\varsigma}
\def\adg{a^{\dag}}

\title{POVM Quantization}
\author{Jean Pierre Gazeau and Barbara Heller}
\address{Laboratoire APC, Universit\'e Paris 7-Denis Diderot, 10, rue A. Domon et L. Duquet, 75205 Paris Cedex13, France}\email{gazeau@apc.univ-paris7.fr}
\address{
Department of Applied Mathematics,
Illinois Institute of Technology,
Chicago, IL 60616}
\email{heller@iit.edu, effe@uchicago.edu}

\date{\today}                                           % Activate to display a given date or no date

\begin{document}
\maketitle
\begin{abstract}
We present a general formalism for giving a measure space paired with a separable Hilbert space  a quantum version based on normalized positive operator-valued measure. The latter are built from families of density operators labelled by points of the measure space. We specially focus on various probabilistic aspects of these constructions. Simple or more elaborate examples  illustrate the procedure: circle, 2-sphere, plane, half-plane. Links with POVM quantum measurement and quantum statistical inference are sketched.
\end{abstract}
\newpage
\tableofcontents

\section{Introduction}
In this paper we propose a quantum analysis, generally non-commutative, of a measure space based on a (normalized) positive operator-valued measure ((N)POVM)\footnote{In order not to spoil the text with too many acronyms, we will keep ``POVM" in our paper to designate a normalized positive operator-valued measure} built from a \textit{density matrix or operator} (in the quantum mechanics terminology) acting on some separable Hilbert state. One key aspect of the procedure is its probabilistic nature. Moreover, beyond the common mathematical language, our approach has or might have some deep connection with quantum measurement based on POVM, quantum probability (see for instance \cite{qp1} with references therein), or quantum statistical inference (see for instance \cite{qsi1} with references therein). Let us just quote from \cite{sosathi13} 
\begin{quote}
\textit{POVMs are the most general measurements one can make on a quantum system and although in principle they can be reformulated as projective measurements on larger spaces, for which filtering results exist, a direct treatment of POVMs is more natural and can simplify the filter computations for some applications.} 
\end{quote}
We also recommend the very clear and concise introduction to the mathematics of quantum physics by  Kuperberg  \cite{kuperberg05}.

Our work lies in the continuation of recent ones concerning what we named \textit{integral quantization} \cite{bergaz14,aagbook13,bercugazro14,wcosmo2,balfrega14} and leading to  applications
shedding a new light on the still problematic question of the relation between classic and quantum worlds. The so-called coherent state (CS) or Berezin or Klauder or anti-Wick or Toeplitz quantizations are particular cases of those integral quantizations of various measure sets.

Our conception of quantization rests upon a trivial observation. We notice that the formalism of classical physics rests upon highly abstract mathematical models, mainly since the invention of infinitesimal calculus, giving us the impression that improbable objects like material phase space points are accessible to measurements. It is true that with an excellent approximation most of the physical phenomenon at our scale can be efficiently apprehended in that way. On the other hand, reasonably realistic  scientists know that such continuous models are highly idealistic and should be viewed so, whatever their powerful predictive qualities. 
Above all, we know that any attempt to maintain our ``classical'' models together with our classical reading of them is not experimentally sustainable over a wide range of phenomenona. A quantization in a certain sense of our \underline{mathematical classical} model (Bohr-Sommerfeld, canonical Dirac, Feynman path integral, geometry, deformation, CS, ...\cite{alienglis05}) is needed to account for observations and predictability. Usually physicists or mathematicians have in mind as a classical structure a phase space or symplectic one which fits with Hamiltonian formalism. In our mind this represents a quite constraining restriction. With our approach, classical mathematical models with minimal structure (like  a measure) might  also be amenable to their quantized versions in our sense.

Now we should answer the natural question ``POVM Quantization for what?''. In quantum physics, the answer is natural and experimentally  justified. Some illuminating examples are given in our previous works \cite{bergaz14,balfrega14} where it has been shown that there is a world  of quantizations leading to equivalent results from a physical point of view \cite{bergayou13}. Starting from general models, not necessarily endowed with some physical flavor, it is interesting to provide a class of noncommutative, ``fuzzy'', versions of them based on normalized POVM and resultant classical probability distributions. The method can be particularly relevant when we have to cope  with geometries presenting singularities, or with subset of manifolds determined by constraints \cite{balfregaCST14}. 

In Section \ref{basquant} we recall the minimal requirements that any  quantization procedure should obey. 
Normalized positive operator-valued measure associated with the triple \textit{measure space}, \textit{Hilbert space}, \textit{density operator}, is presented in  Section \ref{mspovm}.  
The probabilistic content of the formalism is developed in Section \ref{probdenspov}. In Section \ref{cprobdistq} we reverse the approach by asking whether quantum formalism can be directly produced from classical probability theory. 
In Section \ref{CSpovm} we examine the particular case where density operators are rank one, i.e. coherent states projectors. This allows a better understanding of the material introduced in the three previous sections.   
With Section \ref{pintquant} we enter the heart of the subject by explaining in which manner POVM quantization transforms a classical object, function or distribution into a linear  operator in the companion Hilbert space.
In  Section \ref{semclass} semi-classical aspects through lower symbols are examined. 
Covariant POVM quantization based on unitary irreducible representations and relevant Schur's lemma are described  in  Section \ref{covquant}. Then we proceed with more or less elementary illustrations of the method: unit circle (Section \ref{unitcircle}), unit 2-sphere (Section \ref{unitsphere}), plane (Section \ref{plane}), and finally half-plane (Section \ref{halfplane}). 
Some lines for future works and views about  the links with quantum probability  and quantum measurements are sketched in Section \ref{conclusion}. Some necessary material is given in the two appendices.  

\section{Quantization: the basics}
\label{basquant}
First, on a minimal level, we understand quantization
of a  set $X$ and functions on it as a procedure fulfilling
three requirements: linearity, existence of identity and self-adjointness.
More precisely, quantization is: 
\begin{enumerate}
\item A \textbf{linear} map 
\begin{equation}
\label{linmapq}
\mathfrak{Q}:\mathcal{C}(X)\rightarrow\mathcal{A}(\mathcal{H})\,,
\end{equation}
where $\mathcal{C}(X)$ is a vector space of complex-valued functions
$f(x)$ on a set $X$ and $\mathcal{A}(\mathcal{H})$ is a vector
space\footnote{``Vector space'' in a loose sense since the linear superposition of two operators could have an empty domain in infinite-dimensional Hilbert space!} of linear operators 
\begin{equation}
\label{fAf}
\mathfrak{Q}(f)\equiv A_{f}
\end{equation}
in some complex Hilbert space $\mathcal{H}$ such that; 
\item $f=1$ is mapped to the identity operator $I$ on $\mathcal{H}$; 
\item A real function $f$ is mapped to an (essentially) self-adjoint operator
$A_{f}$ in $\mathcal{H}$. 
\end{enumerate}
In a physical or a signal analysis context,
one needs to add structure to $X$ such as measure, topology, manifold
structure, closure under algebraic operations, etc. Besides, one also
has the freedom to  interpret the spectra of classical
$f\in\mathcal{C}(X)$ or quantum $A_{f}\in\mathcal{A}(\mathcal{H})$,
so that they can be chosen as observables (in the terminology used in Physics). And finally, one may add
the requirement of an unambiguous classical limit of the quantum 
quantities, the limit operation being associated with a change of
scale.

\section{POVM for a measure space}
\label{mspovm}
As announced in the introduction, we start from a minimal set of objects:
\begin{itemize}
  \item[(i)] a measure space $\left(X,\mathfrak{B},\nu\right)$ (or $(X,\nu)$ for short), where $\mathfrak{B}$ is the $sigma$-algebra of  $\nu$-measurable subsets, 
  \item[(ii)] a separable Hilbert space $\mathcal{H}$,
  \item [(iii)] an $X$-labeled family of positive semi-definite and unit trace operators (``density matrices or operators'') on $\mathcal{H}$,
 \begin{equation}
\label{famdens}
X\ni x\mapsto\mathsf{\rho}(x)\in\mathcal{L}(\mathcal{H})\,,\quad  \rho(x)>0\, , \quad \mathrm{tr}( \rho(x)) = 1\, , 
\end{equation} 
  and resolving the identity $I$ on $\mathcal{H}$,
\begin{equation}
\label{residrho}
\int_{X}\,\mathsf{\rho}(x)\,\mathrm{d}\nu(x)=I\,,\quad\mbox{in a weak sense}.
\end{equation}
\end{itemize}
 If $X$ is equipped with a suitable topology, then the normalized positive operator-valued measure (POVM) $\mathfrak{m}_{\rho}$
on the corresponding $\sigma$-algebra $\mathfrak{B}_{\rho}(X)$ of Borel sets is defined through the following
map $\Delta$
\begin{equation}
\label{povmap}
\mathcal{B}(X)\ni\Delta\mapsto\mathfrak{m}_{\rho}(\Delta)=\int_{\Delta}\rho(x)\,\mathrm{d}\nu(x)\,.
\end{equation}

\section{Probabilistic density on measure space from POVM }
\label{probdenspov}

There is a straightforward  consequence of the  identity \eqref{residrho} in terms of probability distribution on the original measure space $(X,\nu)$. Given $x_0 \in X$ and applying the corresponding density operator $\rho(x_0)$ on each side of \eqref{residrho} leads to
\begin{equation}
\label{rhorho}
\int_{X}\,\mathsf{\rho}(x_0)\,\mathsf{\rho}(x)\,\mathrm{d}\nu(x)=\rho(x_0)\, . 
\end{equation}
Taking now the trace on each side gives 
\begin{equation}
\label{rhorho}
\int_{X}\,\mathrm{tr}\left(\mathsf{\rho}(x_0)\,\mathsf{\rho}(x)\right)\,\mathrm{d}\nu(x)=\mathrm{tr}\left(\rho(x_0)\right)= 1\, . 
\end{equation}
Hence, the hilbertian formalism combined with the original measure $\nu$ produces the $X$-labelled family of probability distributions 
\begin{equation}
\label{probdist}
X\ni x_0, x \mapsto p_{x_0}(x) = \mathrm{tr}\left(\mathsf{\rho}(x_0)\,\mathsf{\rho}(x)\right)
\end{equation}
on $(X,\nu)$. The nonnegative bounded function $ p_{x_0}(x) \leq 1$ measures in a certain sense the degree of localization of $x$ w.r.t. $x_0$, and vice versa due to the symmetry $ p_{x_0}(x) = p_{x}(x_0) $,
on the measure space $(X,\nu)$. If we consider the particular case where $\rho(x)$ is a rank-one projector operator 
\begin{equation}
\label{rankone}
\rho(x) = |x\rg\lg x|\, , \quad \lg x|x\rg = 1\, , 
\end{equation}
i.e. is  a ``pure coherent state" (see below),
then 
\begin{equation}
\label{rk1inprod}
p_{x_0}(x) = \vert \lg x_0|x\rg\vert^2\, , 
\end{equation}
Thus we could be inclined to introduce the pseudo-distance (triangular inequality is not verified in general)
\begin{align}
\label{psdist1}
\delta(x,x^{\prime}) &:= \left\lbrack-  \ln \frac{\mathrm{tr}(\rho(x)\rho(x^{\prime}))}{\sqrt{\mathrm{tr}((\rho(x))^2)\mathrm{tr}((\rho(x^{\prime}))^2)}}\right\rbrack^{1/2}=\left\lbrack-  \ln \frac{p_{x}(x^{\prime})}{\sqrt{p_{x}(x)\,p_{x^{\prime}}(x^{\prime})}}\right\rbrack^{1/2} \\
\label{psdist2} &=  \delta(x^{\prime},x)\in [0,\infty) \, , \quad \delta(x,x)  = 0\,. 
\end{align}
Note that this quantity becomes infinite as $p_{x}(x^{\prime}) \to 0$. This limit corresponds to orthogonality of vectors $|x\rg$ and $|x^{\prime}\rg$ in the pure CS case. 

Actually, from the fact that any density operator $\rho$ is Hilbert-Schmidt, with norm $\Vert \rho\Vert = \sqrt{\mathrm{tr}\rho \rho^{\dag}} = \sqrt{\mathrm{tr}\rho^2}$, it is exact and could appear as more natural to introduce  the associated distance 
\begin{equation}
\label{distHS}
d_{\mathrm{HS}}(x,x^{\prime}) = \Vert \rho(x) - \rho(x^{\prime})\Vert = \sqrt{\mathrm{tr}(\rho(x) - \rho(x^{\prime}))^2}\, . 
\end{equation}
In reality, this object  forces any pair of points in $X$ to be finitely separated since we have
\begin{equation}
\label{distHSfin}
d_{\mathrm{HS}}(x,x^{\prime}) =  \sqrt{\mathrm{tr}\left((\rho(x))^2  + (\rho(x^{\prime}))^2 - 2\rho(x)\rho(x^{\prime})\right)}\leq  \sqrt{2} \sqrt{1 - \mathrm{tr}(\rho(x)\rho(x^{\prime}))}\leq \sqrt{2}\, . 
\end{equation}
In its general form, a density operator can be written as a statistical mixture of pure states
\begin{equation}
\label{densopmix}
\rho(x) = \sum_{i}p_i(x) |\psi_i(x)\rg\lg \psi_i(x)|\, , \quad \Vert\psi_i(x)\Vert = 1\,, \quad \sum_i p_i(x) = 1\, , \quad 0\leq p_i(x) \leq 1\, . 
\end{equation}
Then the corresponding probability distributions  on $(X,\nu)$ read as 
\begin{equation}
\label{misdist}
p_{x_0}(x) = \sum_{i,j} p_i(x_0) p_j(x) \vert  \lg \psi_i(x_0)|\psi_j(x)\rg\vert^2\, . 
\end{equation}
This can be viewed as the average of the random variable $\vert  \lg \psi_i(x_0)|\psi_j(x)\rg\vert^2 \in [0,1]$  with discrete probability distribution $(i,j) \mapsto p_i(x_0) p_j(x)$.

From the point of view of Bayesian statistical inference,  we may treat $X$ as the ``parameter space of interest", $\nu$ as a probability measure \textit{a priori} on $X$ and then $p_{x_0}(x)$ as a probability density function on $X$, \textit{a posteriori}, given an ``estimated" value  $x_0$ where  $x_0$ derives as a datum from some related random device with probability density function family related to $p_{x_0}(x)$.  Then we would be interested in an associated distance function on $X$ to determine intervals of ``$x-$distance"  around the observed value $x_0$.  Note that for this ``inferred" probability distribution on $X$,  we have a POV measure, not an orthogonal one. From the inference point of view, the inferred probability distribution in this context, in principle, does not have a ``frequency" or ``ensemble" interpretation similarly as is the case for a POV measure. It is the “random experiment” with probability density function related to  $p_{x_0}(x)$ which, in principle, is repeatable and which would derive from a PV measure.

\section{Quantum world from classical probabilistic distribution?}
\label{cprobdistq}
In the previous section, we derived from the ``quantum'' 4-tuple $(X,\nu, \mathcal{H}, \, x\mapsto \rho(x))$ an  $X$-indexed family of ``classical'' probability distributions $p_{x_0}(x) = \mathrm{tr}\left(\mathsf{\rho}(x_0)\,\mathsf{\rho}(x)\right)$. An interesting question then arises: given such a classical family, is it possible to derive a quantum $x\mapsto \rho(x)$? If yes, is there uniqueness? Can we loosely think of quantum formalism as a kind of ``square root'' of classical probability formalism, like quantum spin emerges from ``square roots'' (e.g., Dirac) of scalar wave equations (e.g., Klein-Gordon)?

Let us attempt through a simple example to explore such possibilities. Let $X= \{x_1\, , \, x_2\,,\,, \dotsc, \, x_N\}$ be a finite set equipped with the measure,
\begin{equation}
\label{count}
\int_{X}f(x)\,\ud\nu(x):= \sum_{i=1}^N \nu_i \,f(x_i)\, , \quad \nu_i \geq 0\,. 
\end{equation}
A first observation has to be made concerning the existence of a family of $N$ density matrices $\rho(x_i)$ acting on $\C^n$, i.e. hermitian $n\times n$-matrices with unit trace, which resolve the identity w.r.t. this measure,
\begin{equation}
\label{finresNn}
\sum_{i=1}^N \nu_i \, \rho(x_i) = I\, .
\end{equation}
Taking the trace of each side of this equation yields the constraint on the set of weights $\nu_i$
\begin{equation}
\label{consnu}
\sum_{i=1}^N \nu_i  = n\, .
\end{equation}
To simplify, we suppose that $\nu_i >0$ for all $i$.
In particular, if the measure is uniform, $\nu_i= \nu$ for all $i$, then $\nu = n/N$. 
Another point concerns the cardinal $N$ of $X$ versus the dimension $n$ of $\mathcal{H}$.  
In its full generality, which means in the $n$-rank case, each $n\times n$ density matrix $\rho(x_i)$ is defined by $n -1 + n(n-1)/2 \times 2 = n^2 -1$ real parameters. Moreover, in the present case, these $N$ density matrices are requested to satisfy  the set of equations
issued from \eqref{finresNn}
\begin{equation}
\label{resrhoi}
\sum_{i=1}^N \nu_i \,\rho(x_i)_{ab} = \delta_{ab} \, , \quad 1\leq a\leq  b \leq n\, . 
\end{equation}
Due to \eqref{consnu} they are not independent and represent
$n^2-1$ real constraints. Moreover, these constraints have to be supplemented by the (non trivial!) condition  that, for all $i$, $\rho(x_i)$ is a positive semi-definite matrix. 
This entails that we are left with a maximum of $Nn^2 - N -n^2 + 1= (N-1)(n^2-1)$ free parameters. Hence, as soon as $n\geq 2$, free parameters exist as soon as $N\geq 2$. Let us examine the minimal non trivial case $N= n= 2$. Eq.\;\eqref{finresNn} assumes the $2\times2$ matrix form
\begin{equation}
\label{N2n2}
\nu\,\begin{pmatrix}
   a   &  b  \\
  \bar b    &  1-a
\end{pmatrix}  + (2-\nu)\,\begin{pmatrix}
   a^{\prime}   &  b^{\prime}  \\
  \bar b^{\prime}    &  1-a^{\prime}
\end{pmatrix} = \begin{pmatrix}
   1   &  0\\
  0   &  1
\end{pmatrix}\, , \quad 0\leq \nu \leq 2\, . 
\end{equation} 
This linear relation between two positive matrices implies that they are simultaneous diagonalisable, with respective eigenvalues $0<\lambda\, , \, 1-\lambda<1$, $0<\lambda^{\prime}= (1-\nu\lambda)/(2-\nu)\, , \, 1-\lambda^{\prime}<1$,  with normalized eigenvectors $|e_1\rg$, $|e_2\rg$, forming  an orthonormal basis of $\C^2$. Hence \eqref{N2n2} is just a trivial rewriting of the resolution of the identity in $\C^2$
\begin{equation}
\label{resC2}
(\nu \lambda + (2-\nu) \lambda^{\prime})\,|e_1\rg\lg e_1| + (\nu 1-\lambda^{\prime} + (2-\nu)(1- \lambda^{\prime})\,|e_2\rg\lg e_2|= |e_1\rg\lg e_1| + |e_2\rg\lg e_2| =  I\, .  
\end{equation}

 A second observation is that if all $\rho(x_i)$ are rank one, i.e. $\rho(x_i)= |x_i\rg\lg x_i|$, $\lg x_i|x_i\rg = 1$, then \eqref{finresNn} reads
 \begin{equation}
\label{finresNn1}
\sum_{i=1}^N \nu_i \, |x_i\rg\lg x_i| = I\, 
\end{equation}
which means that the set $\{ \sqrt{\nu_i}\, |x_i\rg\}$ is a Parseval frame \cite{benedetto03,han07,cotgaz10,cotgavour11}. Such an identity is possible if $N\geq n$, and if $N=n$, then $\nu_i = 1$ for all $i$ and $\{ |x_i\rg\}$ is an orthonormal basis. 

Suppose that a family $p_{ij}= p_{x_i}(x_j)= p_{x_j}(x_i)$ of $N$ probability distributions is defined on the measure space $(X,\nu)$, i.e. a set of $N(N+1)/2$   non-negative numbers $p_{ij}= p_{ji}$ obeying
\begin{equation}
\label{disprobN}
\sum_{j=1}^N \nu_j \,p_{ij} = 1\, , \quad i=1,2,\dotsc, N\,.  
\end{equation}
So we are left with $N(N+1)/2 -N= N(N-1)/2$ free parameters. 
Inspired by \eqref{probdist}, we attempt to determine  a set of $N$ density matrices $\rho(x_i)$ from the following identities
\begin{equation}
\label{relprho}
 \mathrm{tr}\left(\mathsf{\rho}(x_i)\,\mathsf{\rho}(x_j)\right)= p_{ij}= p_{x_i}(x_j)= p_{x_j}(x_i)\,. 
\end{equation}

Now, \eqref{relprho} leads to the set of $N + N(N-1)/2= N(N+1)/2$  real quadratic equations 
\begin{equation}
\label{deteq1}
p_{ij}= \sum_{1\leq a \leq n} \rho(x_i)_{aa}\,\overline{\rho(x_j)_{aa}}  + 2\mathrm{Re}\sum_{1\leq a < b\leq n} \rho(x_i)_{ab}\,\overline{\rho(x_j)_{ab}}\, ,
\end{equation}
Actually there are not independent since, for each $i$, applying $\sum_{j=1}^N \nu_j$ on each side gives 1. So $N(N-1)/2$ of these equations are independent. It follows the necessary condition
\begin{equation}
\label{condNn}
N(N-1)/2 \leq N n^2 -N -n^2 +1 \Leftrightarrow N^2-  N (2n^2-1) + 2n^2 -2 \leq 0
\end{equation} 
for having nontrivial solutions, and uniqueness might hold with $N^2-  N (2n^2-1) + 2n^2 -2= (N-1)(N-2n^2+2)= 0$. Hence, Condition \eqref{condNn} defines the allowed range for $N$ with respect to $n$
\begin{equation}
\label{allowr}
1\leq N\leq 2n^2-2\, . 
\end{equation}
%It is instructive to go through the minimal nontrivial case $n= 2$ and $N= 2n^2-1= 6$. This case is examined in Appendix \ref{N6n2}. 
%Another way to grasp the problem is to start from a metric defined on $X$ and to use the relation \eqref{distHS}. Suppose that a distance $d(x_i,x_j)$ on $X$ is defined through a set $\{d_{ij}\}$, $i>j$,  of $N/(N-1)/2$ positive and bounded numbers, $d(x_i,x_j) = d_{ij}$, i.e., is determined by  the symmetric matrix $D= \left(d_{ij}\right)$ with null diagonal elements, $0< d_{ij} = d_{ji}< \infty$ for $i\neq j$, $d_{ii}= 0$, and $ d_{ik} \leq  d_{ij}+  d_{jk}$. Is it possible to derive from the following $N(N-1)/2$  quadratic equations
%\begin{equation}
%\label{deteq2}
%d^2_{ij}= \mathrm{cst}\times\sum_{1\leq a \leq b\leq n}\left(\rho(x_i)_{ab}- \rho(x_j)_{ab}\right)^2\, .
%\end{equation}
%for their respective matrix elements an unknown  set of $N$ density matrices $\rho(x_i)$ in $\C^n$. Since these equations involve also $n^2-1$ unknown real numbers, a complete determination is possible if the number $N$ obeys the inequality
%\eqref{detineq}.
On the other hand, in the minimal case corresponding to rank-one density matrices $\rho(x_i) = |x_i\rg\lg x_i|$, i.e. coherent states, the probabilities are given by 
\begin{equation}
\label{pijxixj}
p_{ij} = \mathrm{tr}(\rho(x_i)\rho(x_j))= \vert \lg x_i|x_j\rg \vert^2 :=  \cos^2 (\theta_{ij})\, .
\end{equation}
Hence, these probabilities must obey the $N$ constraints $p_{ii}= 1$ to be added to the $N$ ones \eqref{disprobN}. This means we are left with that $N(N-1)/2 -N = N(N-3)/2$ free parameters. Let us now express the resolution of the identity \eqref{finresNn1}. In terms of the respective coordinates $\xi_{li}$ of vectors $|x_i\rg$ with respect to an orthonormal basis $\{ |e_l\rg\}$ in $\C^n$.
\begin{equation}
\label{rescsN}
\sum_{i=1}^N\nu_i |x_i\rg\lg x_i|= \sum_{l,l^{\prime} = 1}^n \left\lbrack \sum_{i=1}^N \nu_i \,\xi_{li}\, \overline{\xi_{l^{\prime}i}}\right\rbrack |e_l\rg\lg e_{l^{\prime}}| = I \Leftrightarrow \sum_{i=1}^N \nu_i \,\xi_{li}\, \overline{\xi_{l^{\prime}i}}= \delta_{ll^{\prime}}\, . 
\end{equation}
Now, each projector $\rho(x_i) = |x_i\rg\lg x_i|$ is  defined \textit{a priori} by  $2n-2$ real coordinates (one constraint  is for normalization, $\mathrm{tr}(|x_i\rg\lg x_i|) = \lg x_i|x_i\rg = 1$, the other one being  for arbitrary phase). There are $N$ such projectors, so there are $2N(n-1)$ real parameters. From \eqref{rescsN} the latter are submitted to  
\begin{itemize}
  \item $n-1$ independent real constraints issued from the diagonal $l=l^{\prime}$,
  \item $n(n-1)$ real independent constraints  issued from the off-diagonals $l\neq l^{\prime}$ 
\end{itemize}
Hence, like in \eqref{condNn}, we obtain the necessary condition
\begin{equation}
\label{condNncs}
N(N-1)/2 -N \leq 2N(n-1) -n^2 +1\Leftrightarrow N^2 -N(4n-1) + 2n^2 -n \leq 0
\end{equation} 
for having nontrivial solutions, and uniqueness (up to $n$ phases) might hold with $N^2 -N(4n-1) + 2n^2 -n= 0$. This is possible for $N$ in the range
\begin{equation}
\label{allrangeNCS}
\max\left(n,\frac{1}{2}\left[ 4n-1 - \sqrt{8n^2 -8n +9}\right] \right) < N \leq \frac{1}{2}\left[ 4n-1 - \sqrt{8n^2 -8n +9}\right] \, . 
\end{equation}

\section{POVM from coherent states}
\label{CSpovm}
In this section we describe a simple method \cite{gazeaubook09} for obtaining  coherent states $|x\rg$ such that $\rho(x) = |x\rg\lg x|$. We start from another measure space $(X,\mu)$ and consider 
 the Hilbert space $L^2(X,\mu)$ of complex square integrable functions on $X$ with respect to the measure $\mu$. One then chooses  in it an orthonormal set 
$\mathcal{O}$ of functions $\phi_n(x)$ (set aside the question of evaluation map in their respective equivalence classes), satisfying the finiteness and positiveness conditions  
\begin{equation}
\label{kercondCS}
0< \mathcal{N}(x) := \sum_n \vert\phi_n(x)\vert^2 < \infty\quad \mbox{(a.e.)}
\end{equation}
 and in one-to-one correspondence  with the elements of an orthonormal basis $\{  | e_n\rg \}$ of the Hilbert space $\mathfrak{H}$
\begin{equation}
\label{corrfinen}
 | e_n\rg \leftrightarrow  \phi_n\, . 
\end{equation} 
 There results a family $\mathcal{C}$ of unit vectors $|x\rangle$, the coherent states, in $\mathfrak{H}$, which are labelled by elements of $X$ and which resolve the identity operator in $\mathfrak{H}$ with respect to the measure
 \begin{equation}
\label{mesmunu}
\ud \nu(x) = \mathcal{N}(x)\, \ud\mu(x)\, , 
\end{equation} 
\begin{equation}
\label{defcs}
X\ni x \mapsto | x\rg =\frac{1}{\sqrt{\mathcal{N}(x)}}\sum_{n}\overline{\phi_n(x)} | e_n\rg\in \mathfrak{H}\, . 
\end{equation}
\begin{equation}
\label{resun}
\lg x| x\rg = 1\, , \quad \int_{X} \, | x\rg\lg x| \, \mathcal{N}(x)\, \ud\mu(x)= \int_{X} \, | x\rg\lg x| \, \ud\nu(x) = I\, . 
\end{equation}
This certainly represents the most straightforward way to build  total families of states resolving the identity in $\mathfrak{H}$. Underlying the construction, there is a Bayesian content \cite{algahe08}, based or not on experimental evidences or on selective information choice, namely,  an interplay between the set of probability distributions 
\begin{equation}
\label{contprobbay}
x \mapsto \vert \phi_n(x)\vert^2 \quad \mbox{from} \quad \int_X \, \vert \phi_n(x)\vert^2\, \ud\mu(x) =1\, , 
\end{equation}
labelled by $n$,  on the classical measure space $(X,\mu)$, and the discrete set of probability distributions 
\begin{equation}
\label{disprobbay}
n \mapsto \vert \phi_n(x)\vert^2/\mathcal{N}(x) \quad \mbox{from} \quad \mathcal{N}(x)= \sum_n \vert\phi_n(x)\vert^2\, .
\end{equation}
In this CS case, the probability distribution  
 \begin{equation}
\label{rk1inprodCS}
p_{x_0}(x) = \vert \lg x_0|x\rg\vert^2= \vert {\sf K} (x_0,x)\vert^2\, , 
\end{equation}
 is expressed in terms of the reproducing kernel ${\sf K}$ w.r.t. the measure $\ud \nu(x)$
 \begin{equation}
\label{repkern}
{\sf K}(x,x^{\prime}) = \lg x|x^{\prime}\rg \frac{1}{\sqrt{\mathcal{N}(x)\,\mathcal{N}(x^{\prime})}}\sum_{n,n^{\prime}} \phi_n(x)\overline{\phi_{n^{\prime}}(x)}\,. 
\end{equation}

\section{POVM integral quantization}
\label{pintquant}

With the above material at hand, the integral quantization of complex-valued functions $f(x)\in\mathcal{C}(X)$
is formally defined as the linear map
\begin{equation}
\label{povmquantf}
f\mapsto A_{f}=\int_{X}\, f(x)\,\rho(x)\,\mathrm{d}\nu(x)\,.
\end{equation}
This map is properly defined if the operator $A_{f}\in\mathcal{A}(\mathcal{H})$
is understood as the sesquilinear form 
\begin{equation}
\label{seqAf}
B_{f}(\psi_{1},\psi_{2})=\int_{X}f(x)\,\langle\psi_{1}|\rho(x)|\psi_{2}\rangle\,\mathrm{d}\nu(x)\,,
\end{equation}
defined on a dense subspace of $\mathcal{H}$. If $f$ is real and
at least semi-bounded and since $\rho\left(x\right)$ is positive, the Friedrich's
extension \cite{reedsimon2} of $B_{f}$ univocally defines a self-adjoint
operator. If $f$ is not semi-bounded, there is no natural choice
of a self-adjoint operator associated with $B_{f}$.
In this last case, in order to construct $B_{f}$ as an observable,
we need to know more about the space of states $\mathcal{H}$ in order to examine the existence of self-adjoint extensions (e.g. boundary conditions in the case of domains defined for wave functions).

Note that the above quantization may be extended to objects which are more general than functions. We think of course to distributions if relevant  structure of $X$ allows to properly define them.  Suppose that the measure set $(X,\nu)$ is also a smooth manifold of dimension $n$, on which is defined the space $\mathcal{D}'(X)$ of distributions  as the topological dual of the (LF)-space $\Omega_c^n(X)$ of compactly supported $n$-forms on $X$ \cite{grosser08}. Some of these distributions, e.g.  $\delta(u(x))$, express geometrical constraints. Extending the map \eqref{povmquantf} yields the quantum version $A_{\delta(u(x))}$ of these constraints. 

A different starting point for quantizing constraints, more in Dirac's spirit \cite{dirac64} 
would consist in quantizing the function $u \mapsto A_u$ and determining the kernel of the operator $A_u$. Both methods are obviously not equivalent, except for a few cases. This question of equivalence/difference gives rise to controversial opinions in fields like quantum gravity or quantum cosmology. Elementary examples illustrating this difference are worked out  in \cite{balfrega14}.

\section{Semi-classical aspects and quantum measurement through lower symbols}
\label{semclass}

We arrive at the point where the probability distribution \eqref{probdist} makes sense in regard to the objects $f$ (functions or more singular entities) to be quantized. Indeed, some of the properties (if not all!) of the operator $A_f$ can be grasped by examining the 
function $\check{f}(x)$ defined as 
\begin{equation}
\label{lowsymbpovm}
A_{f}\mapsto\check{f}(x):= \mathrm{tr}(\rho(x)\,A_f)\,,
\end{equation}
and named, within the context of Berezin quantization \cite{berezin74}, lower (Lieb) or covariant (Berezin) symbol. 
Now, this quantity represents the local averaging of the original $f$ with respect to the probability distribution \eqref{probdist}
\begin{equation}
\label{lowsymbmap}
f(x) \mapsto \check{f}(x)= \int_{X}f(x^{\prime})\,\mathrm{tr}(\rho(x)\rho(x^{\prime}))\,\mathrm{d}\nu(x^{\prime})= \int_{X}f(x^{\prime})\,p_x(x^{\prime})\,\mathrm{d}\nu(x^{\prime})\, . 
\end{equation} 
This construction is a generalization
of the so-called Bargmann-Segal transform (see for instance \cite{stenzel94,hall94}). 
Besides, from functional properties of the lower symbol $\check{f}$
one may investigate certain quantum features, such as, e.g., spectral
properties of $A_{f}$. Also, the map \eqref{lowsymbmap} represents in general a regularization of the original, possibly extremely singular, $f$. 
Another point deserves to be mentioned here. It concerns the analogy of the present formalism with quantum measurement.
In a quantum physics context for which $A_f$ is a self-adjoint operator or observable of a system, and given a density operator  $\rho_m= \sum_i q_i  |\phi_i\rg\lg\phi_i|$ describing  the mixed state of  an ensemble such that each of the pure states $ |\phi_i\rg$ occurs with probability $q_i$, the expectation value of the measurement  is given by 
\begin{equation}
\label{measexpect}
\mathrm{tr}\left(\rho_m A_f\right) =  \int_{X}f(x)\,\mathrm{tr}(\rho_m\rho(x))\,\mathrm{d}\nu(x)\, . 
\end{equation}
Hence, it can be also viewed as the average of the original $f$ with respect to the probability density 
\begin{equation}
\label{measprobdens}
p_m(x):= \mathrm{tr}(\rho_m\rho(x))\, . 
\end{equation}
Of course, this $\rho_m$ can be one element $\rho_m = \rho(x_0)$ of the family of density operators from which is issued the considered quantization. 
Inspired by  ideas developed during the two last decades by  various authors, particularly  Busch, Grabowski, and Lahti in  “Operational Quantum Physics”\cite{bugrala95}, and Holevo in ``Probabilistic
and Statistical Aspects of Quantum Theory"\cite{holevo11}, we turn our attention to classical ``smeared'' form such as described in these books. If one validates the assumption that any quantum observable is issued from our POVM quantization procedure, then its measurement can be expressed as in \eqref{measexpect}. This should shed a new classical  light on the quantum perspective, since the usual integral representation of $\mathrm{tr}\left(\rho_m A_f\right)$ is issued from the spectral decomposition of the self-adjoint $A_f$ with spectral measure $\mathrm{d}E (\lambda)$:
\begin{equation}
\label{measspecdec}
\mathrm{tr}\left(\rho_m A_f\right) =  \int_{\R}\lambda\,\mathrm{tr}(\rho_m\,\mathrm{d}E (\lambda))\, .
\end{equation}
We point out the ``circular" nature of our procedure.  On the one hand, we use POVM to quantize classical functions. On the other hand, we obtain a POVM quantum measurement,  interpreted as an inverse transform yielding a ``semi-classical object"  which, in the statistical inference context, yields an inferred probability distribution.  In that sense, we treat quantization and measurement as two aspects of the same construct. 

\section{Covariant POVM quantizations}
\label{covquant}

In explicit constructions of  density operator families and related POVM  quantization, the theory of Lie group representations offers a wide range of possibilities. Let $G$ be a Lie group with left Haar measure $\mathrm{d}\mu(g)$,
and let $g\mapsto U\left(g\right)$ be a unitary irreducible representation
(UIR) of $G$ in a Hilbert space $\mathcal{H}$. Pick a density
operator $\rho$ on $\mathcal{H}$ and let us transport it under representation operators $U(g)$. Its orbit is the family of density operators
\begin{equation}
\label{orbitrho}
\rho\left(g\right):=U\left(g\right)\,\rho\, U^{\dagger}\left(g\right)\,, \quad \rho\left(e\right) = \rho\, .
\end{equation}
 Suppose that the operator
\begin{equation}
\label{intrhoR}
R:=\int_{G}\rho\left(g\right)\mathrm{d}\mu\left(g\right)\,, 
\end{equation}
is defined in a weak sense. From the left invariance of $\mathrm{d}\mu(g)$
we have 
\begin{equation}
\label{comR}
U\left(g_{0}\right)\,R\,U^{\dagger}\left(g_{0}\right)=\int_{G}\rho\left(g_{0}g\right)\mathrm{d}\mu\left(g\right)=R\,,
\end{equation}
and so $R$ commutes with all operators $U(g)$, $g\in G$. Thus,
from Schur's Lemma, $R=c_{\rho}I$ with 
\begin{equation}
\label{crho}
c_{\rho}=\int_{G}\mathrm{tr}\left(\rho_{0}\rho\left(g\right)\right)\mathrm{d}\mu\left(g\right)\,,
\end{equation}
where the density operator $\rho_{0}$ is chosen in order
to make the integral converge. This family of operators provides the
following resolution of the identity 
\begin{equation}
\int_{G}\rho\left(g\right)\mathrm{d}\nu\left(g\right)=I,\quad  \mathrm{d}\nu\left(g\right):=\frac{\mathrm{d}\mu\left(g\right)}{c_{\rho}}\,.\label{eq:resolution}
\end{equation}
Let us examine in more detail the above procedure in the case of square
integrable UIR's (e.g. affine group, see below). For a square-integrable UIR $U$
for which $\left\vert \eta\right\rangle $ is an admissible unit vector,
i.e., 
\begin{equation}
\label{admissceta}
c(\eta):=\int_{G}\mathrm{d}\mu(g)\,|\left\langle \eta\right\vert U\left(g\right)\left\vert \eta\right\rangle |^{2}<\infty\,,
\end{equation}
the resolution of the identity is obeyed by the family of coherent states for the group $G$
\begin{equation}
\label{GCSresunit}
\left\vert \eta_{g}\right\rangle \left\langle \eta_{g}\right\vert =\rho\left(g\right)\ ,\ \rho:=\left\vert \eta\right\rangle \left\langle \eta\right\vert \,,\ \left\vert \eta_{g}\right\rangle =U(g)\left\vert \eta\right\rangle \,.
\end{equation}
This property is easily extended to square-integrable UIR $U$  for which $\rho$ is an ``admissible'' density operator, $c(\eta)= \int_G \ud\mu(g) \, \vert \mathrm{tr} \rho U(g)\vert^2 < \infty$.   
 Resolution of the identity then is obeyed by the family: $\rho(g)= U(g) \rho U^{\dag}(g)$ 

This allows an integral quantization of complex-valued functions on
the group 
\begin{equation}
\label{Gcovquant}
f\mapsto A_{f}=\int_{G}\,\rho(g)\, f(g)\mathrm{d}\nu(g)\,,
\end{equation}
which is covariant in the sense that 
\begin{equation}
\label{Gcovprop}
U(g)A_{f}U^{\dagger}(g)=A_{U_{r}(g)f}\,.
\end{equation}
In the case when $f\in L^{2}(G,\mathrm{d}\mu(g))$, the quantity $(U_{r}(g)f)(g^{\prime}):=f(g^{-1}g^{\prime})$
is the regular representation. From the lower symbol we obtain a generalization
of the Berezin or heat kernel transform on $G$ 
\begin{equation}
\label{Gberheat}
\check{f}(g):=\int_{G}\, tr(\rho(g)\,\rho(g^{\prime}))\, f(g^{\prime})\mathrm{d}\nu(g^{\prime})\,.
\end{equation}
In the absence of square-integrability over $G$, there exists a definition
of square-integrable covariant coherent states with respect to a left
coset manifold $X=G/H$, with $H$ a closed subgroup of $G$, equipped
with a quasi-invariant measure $\nu$ \cite{aagbook13}.

\section{The example of the unit circle}
\label{unitcircle}
We start our series  of examples with one of the most elementary ones. Actually it  is rich both in fundamental aspects and pedagogical resources. The measure set is the unit circle equipped with its uniform (Lebesgue) measure:
\begin{equation}
\label{measS1}
X= \mathbb{S}^1\, , \quad \mathrm{d}\nu(x) = \frac{\ud \theta}{\pi}\, , \quad \theta \in [0, 2\pi)\, .
\end{equation}
The Hilbert space is the euclidean plane $\mathcal{H}= \R^2$. The group $G$ is the group SO(2) of rotations in the plane. As described at length in Appendix \ref{pararho}, the most general form of a real density matrix can be given, as a $\pi$-periodic matrix, in terms of the polar coordinates $(r,\phi)$ of a point in the unit disk:
\begin{equation}
\label{standrhomain}
\rho_{r,\phi}= \begin{pmatrix}
  \frac{1}{2}  + \frac{r}{2}\cos2\phi  &   \frac{r}{2}\sin2\phi  \\
\frac{r}{2}\sin2\phi    &   \frac{1}{2}  - \frac{r}{2}\cos2\phi
\end{pmatrix}= \rho_{r,\phi+\pi}\, , \quad 0\leq r\leq 1\, , \ 0\leq \phi < \pi\, . 
\end{equation}
We notice that for  $r=1$  the density matrix is just the orthogonal projector on the unit vector $|\phi\rg$ with polar angle $\phi$:
\begin{equation}
\label{standrhomain1}
\rho_{1,\phi}= \begin{pmatrix}
\cos^2\phi  &   \cos\phi \,\sin\phi \\
 \cos\phi \,\sin\phi &  \sin^2\phi 
\end{pmatrix}= |\phi\rg\lg \phi|= |\phi + \pi\rg\lg \phi + \pi|\, . 
\end{equation}
Due to the covariance property \eqref{rotrho}, we define the family of density operators 
\begin{equation}
\label{rotrhomain}
 \rho_{r,\phi}(\theta)= \mathcal{R}\left(\theta\right) \rho_{r,\phi}\mathcal{R}\left(-\theta\right)=  \rho_{r,\phi+\theta}\, , \quad 0\leq \theta< 2\pi\, . 
\end{equation}
where the rotation matrix $\mathcal{R}(\theta) $ is defined by \eqref{rotmat}.  This family resolves the identity
\begin{equation}
\label{margomegamain}
\int_0^{2\pi} \rho_{r,\phi}(\theta) \, \frac{\ud\theta}{\pi}= I\,.
\end{equation}
It follows the $\mathbb{S}^1$-labelled family of probability distributions on $(\mathbb{S}^1\,, \, \ud \theta/\pi)$
\begin{equation}
\label{probdistcirc}
 p_{\theta_0}(\theta) = \mathrm{tr}\left(\mathsf{\rho_{r,\phi}}(\theta_0)\,\mathsf{\rho_{r,\phi}}(\theta)\right)= \frac{1}{2}\left(1 + r^2\cos 2(\theta-\theta_0)\right)\, . 
\end{equation}
At $r= 0$ we get the uniform probability on the circle whereas at $r=1$ we get the ``pure state'' probability distribution
\begin{equation}
\label{purdistcirc}
p_{\theta_0}(\theta) = \cos^2\left(\theta-\theta_0\right)\, . 
\end{equation}
Hence, the parameter $r$ can be thought as the inverse of a ``noise'' temperature $r\propto 1/T$. 
The pseudo-distance on $\mathbb{S}^1$ associated with \eqref{probdistcirc} is given by 
\begin{equation}
\label{psdistS1}
\delta^2_r(\theta,\theta^{\prime}) = -\ln\frac{1 + r^2\cos 2(\theta-\theta^{\prime})}{1+r^2}\, ,
\end{equation}
which reduces at small $\theta-\theta^{\prime}$ to 
\begin{equation}
\label{psdistSsm}
\delta_r(\theta,\theta^{\prime}) \approx \frac{2r}{\sqrt{1+r^2}}\,\vert\theta-\theta^{\prime}\vert\, .
\end{equation}
On the other hand, the distance $d_{\mathrm{HS}}$  defined by \eqref{distHS} reads in the present case
\begin{equation}
\label{distHSS1}
d_{r;\mathrm{HS}}(\theta,\theta^{\prime}) = \sqrt{\mathrm{tr}(\rho_{r,\phi}(\theta) - \rho_{r,\phi}(\theta^{\prime}))^2}= \sqrt{2}r\vert \sin(\theta-\theta^{\prime})\vert\, 
\end{equation}
which reduces at small $\theta-\theta^{\prime}$ to \eqref{psdistSsm} up to a constant factor. 

The quantization of a function (or distribution) $f(\theta)$ on the circle based on \eqref{margomegamain}   leads to 
the 2$\times$2 matrix operator
\begin{equation}
\label{qtfrhor}
f \mapsto A_f = \int_0^{2\pi} f(\theta) \rho_{r,\phi}(\theta) \, \frac{\ud\theta}{\pi}= \begin{pmatrix}
  \lg f\rg  + \frac{r}{2}C_c\left(R_{-\phi}f\right)  &   \frac{r}{2}C_s\left(R_{-\phi}f\right) \\
\frac{r}{2}C_s\left(R_{-\phi}f\right)   &   \lg f\rg - \frac{r}{2}C_c\left(R_{-\phi}f\right)
\end{pmatrix}\,,
\end{equation}
where $ \lg f\rg:= \frac{1}{2\pi}\int_0^{2\pi}f(\theta)\,\ud\theta$ is the average of $f$ on the unit circle and $R_{\phi}(f)(\theta) := f(\theta-\phi)$. The symbols $C_c$ and $C_s$ are for the cosine and sine doubled angle Fourier coefficients of $f$,
\begin{equation}
\label{CcCs}
C_c(f) = \int_0^{2\pi} f(\theta) \cos2\theta \, \frac{\ud\theta}{\pi}\, , \quad C_s(f) = \int_0^{2\pi} f(\theta) \sin2\theta \, \frac{\ud\theta}{\pi}\, . 
\end{equation}
The simplest function to be quantized is the angle function $\gimel(\theta)$, i.e. the  $2\pi$-periodic extension of $\gimel(\theta)  = \theta$ for $\theta \in [0,2\pi)$,
\begin{equation}
\label{qtfrhor}
 A_{\gimel}= \begin{pmatrix}
  \pi  +  \frac{r}{2}\sin 2\phi  &   - \frac{r}{2}\cos2\phi \\
- \frac{r}{2}\cos2\phi   &  \pi -  \frac{r}{2}\sin2\phi
\end{pmatrix}\,.
\end{equation}
 Its eigenvalues are $\pi   \pm \dfrac{r}{2}$ with corresponding  eigenvectors $\left|\phi \mp\dfrac{\pi}{4}\right\rg$. 
Its lower symbol is given by the smooth function
\begin{equation}
\label{lowsagluc}
\check{\gimel}(\theta) = \pi - r^2\sin\theta\,. 
\end{equation}

\section{The example of the unit 2-sphere}
\label{unitsphere}
The measure set is  the unit sphere equipped with its rotationally invariant  measure:
\begin{equation}
\label{measS2}
X= \mathbb{S}^2\, , \quad \mathrm{d}\nu(x) = \frac{\sin\theta\,\ud \theta\, \ud \phi}{2\pi}\, , \quad \theta \in [0, \pi]\, , \quad \phi\in [0,2\pi)\, .
\end{equation}
The Hilbert space is now  $\mathcal{H}= \C^2$.  The group $G$ is the group SU(2) of $2\times2$-unitary matrices with determinant 1. We give in Appendix \ref{allSU2} the essential about notations and relations with quaternions. 

The unit ball $\mathbb{B}$ in $\mathbb{R}^{3}$ parametrizes the
set of $2\times2$ complex density matrices $\rho$. Indeed, given
a 3-vector $\vec{d}\in\mathbb{R}^{3}$ such that $\Vert\vec{d}\Vert\leq1$,
a general density matrix $\rho$ can be written as 
\begin{equation}
\label{densrhod}
\rho\equiv \rho_{\vec{d}}= \frac{1}{2}(1-i\,\underset{\backsim}{\mathbf{d}})\,.
\end{equation}
 We have used for convenience the  quaternionic representation $\vec{d}\equiv (0,\underset{\backsim}{\mathbf{d}})\in \mathbb{H}$ of the vector $\vec{d}\in \R^3$ (see Appendix \ref{sec:su2} for notations and formulas).
If $\Vert\vec{d}\Vert=1$, i.e. $\vec{d}\in S^{2}$ (``Bloch sphere\textquotedblright{}\ in
this context), with spherical coordinates $(\theta,\phi)$, then $\rho$
is the pure state 
\begin{equation}
\label{purstarho}
\rho=\left\vert \theta,\phi\right\rangle \left\langle \theta,\phi\right\vert \,.
\end{equation}
Note that the above column vector has to be viewed as the spin $j=1/2$
coherent state in the Hermitian space $\mathbb{C}^{2}$ with orthonormal
basis $\left\vert j=1/2,m=\pm1/2\right\rangle $: 
\begin{equation}
\label{spinstate}
\left\vert \theta,\phi\right\rangle =\cos\frac{\theta}{2}\left\vert \frac{1}{2},\frac{1}{2}\right\rangle +\sin\frac{\theta}{2}e^{i\phi}\left\vert \frac{1}{2},-\frac{1}{2}\right\rangle \,.
\end{equation}
Let us now transport the density matrix $\rho$
by using the two-dimensional complex representation of rotations in
space, namely the matrix SU(2) representation.  For $\xi\in SU\left(2\right)$, one defines
the family of density matrices labelled by $\xi$: 
\begin{equation}
\label{rhorxi}
\rho_{\vec{d}}(\xi):=\xi\rho\bar{\xi}=\frac{1}{2}(1-i\xi\underset{\backsim}{\mathbf{d}}\bar{\xi})\,.
\end{equation}
In order to get a one-to-one correspondence with the points of the 2-sphere, we restrict the elements of SU(2) to 
those corresponding to the rotation $\mathcal{R}_{\theta,\phi}$ bringing the unit vector $\hat k$ pointing to the North pole  to the vector with spherical coordinates $(\theta,\phi)$, as described in \eqref{rotspec},
\begin{equation}
\label{rotspecmain}
\rho_{\vec{d}}(\theta,\phi):= \xi\left(\mathcal{R}_{\theta,\phi}\right)\,\rho_{\vec{d}}\,\bar\xi\left(\mathcal{R}_{\theta,\phi}\right)\, , 
\end{equation}
with
\begin{equation}
\label{xiRthphi}
\xi\left(\mathcal{R}_{\theta,\phi}\right)= \left(\cos\frac{\theta}{2}\,, \,\sin\frac{\theta}{2} \hat u_{\phi}\right)\, , \  \hat u_{\phi}= (-\sin\phi,\cos\phi,0)\, . 
\end{equation}
The value of the integral for $\vec{r}=(x,y,z)$
\begin{equation}
\int_{\mathbb{S}^2} \rho_{\vec{d}}(\theta,\phi)\,\frac{\sin\theta\,\ud \theta\, \ud \phi}{2\pi}= \begin{pmatrix}
  1    & x+iy   \\
 x-iy     &  1
\end{pmatrix}\,.\label{resmatS2}
\end{equation}
shows that the resolution of the unity is achieved with $\vec{d}= d\,\hat k$, $0\leq d \leq 1$ only. Then, it is clear that 
\begin{equation}
\label{dkrotr}
\rho_{d \hat k }(\theta,\phi) = \rho_{\vec{r}}= \frac{1}{2}\begin{pmatrix}
1+ r\cos\theta      &  r\sin\theta\,e^{i\phi}  \\
   r\sin\theta\,e^{-i\phi}     &  1- r\cos\theta 
\end{pmatrix}\, , \quad d = \Vert \vec r\Vert \equiv r\,.  
\end{equation}
It is with this strong restriction and the simplified notation 
\begin{equation}
\label{simplnot}
 \rho_{d\hat k}(\theta,\phi)\equiv \rho_r(\theta,\phi)
\end{equation}
that we go forward to the next calculations with the resolution of the unity 
\begin{equation}
\label{S2resun}
\int_{\mathbb{S}^2} \rho_{r}(\theta,\phi)\,\frac{\sin\theta\,\ud \theta\, \ud \phi}{2\pi}= I\,. 
\end{equation} 
Note that  the resolution of the identity with the SU(2) transport of  a generic density operator \eqref{densrhod} is possible only if we integrate on the whole group, as it was done in \cite{balfrega14}.

The $\mathbb{S}^2$-labelled family of probability distributions on $(\mathbb{S}^2\,, \, \sin\theta\,\ud \theta\, \ud \phi/2\pi)$
\begin{align}
\label{probdisph}
 \nonumber p_{\theta_0,\phi_0}(\theta,\phi) &= \mathrm{tr}\left(\mathsf{\rho_{r}}(\theta_0,\phi_0)\,\mathsf{\rho_{r}}(\theta,\phi)\right)= \frac{1}{2}\left(1 + r^2 \hat r_0\cdot \hat r)\right)\\
 &= \frac{1}{2}\left(1 + r^2(\cos\theta_0\cos\theta + \sin\theta_0\sin\theta\cos(\phi_0-\phi) \right)\, . 
\end{align}
At $r= 0$ we get the uniform probability on the sphere whereas at $r=1$ we get the probability distribution corresponding to the spin 1/2 CS \eqref{spinstate},
\begin{equation}
\label{purdistsph}
p_{\theta_0,\phi_0}(\theta,\phi) = \vert \lg \theta_0,\phi_0 | \theta,\phi\rg \vert^2 \, . 
\end{equation}
Like for the unit circle, the parameter $r$ can be viewed as the inverse of a ``noise'' temperature $r\propto 1/T$. 

The pseudo-distance on $\mathbb{S}^2$ associated with \eqref{probdisph} is given by 
\begin{equation}
\label{psdistS2}
\delta^2_r\left((\theta, \phi)\,,(\theta^{\prime},\phi^{\prime})\right) = -\ln\frac{1 + r^2\left(\cos\theta\cos\theta^{\prime} + \sin\theta\sin\theta^{\prime}\cos(\phi-\phi^{\prime})\right)}{1+r^2}\, ,
\end{equation}
which reduces at small $\theta-\theta^{\prime}$ and  $(\phi^{\prime}-\phi)$ to 
\begin{equation}
\label{psdistS2sm}
\delta_r\left((\theta, \phi)\,,(\theta^{\prime},\phi^{\prime})\right) \approx \frac{r}{\sqrt{1+r^2}}\,\sqrt{(\theta-\theta^{\prime})^2 + (\phi-\phi^{\prime})^2\sin^2\left(\frac{\theta + \theta^{\prime}}{2}\right)}\, .
\end{equation}
The distance $d_{\mathrm{HS}}$   reads
\begin{equation}
\label{distHSS2}
d_{r;\mathrm{HS}}(\theta,\theta^{\prime}) = \sqrt{\mathrm{tr}(\rho_r(\theta,\phi) - \rho_r(\theta^{\prime},\phi^{\prime}))^2}= \frac{r}{\sqrt{2}}\Vert \hat r - \hat{r}^{\prime}\Vert= \frac{1}{\sqrt{2}}\Vert \vec r - \vec{r}\mspace{2mu}^{\prime}\Vert\,, 
\end{equation}
which is the usual distance on the sphere with radius $r$ issued from the euclidean one. 
which reduces at small $\theta-\theta^{\prime}$ to \eqref{psdistSsm} up to a constant factor. 
The quantization of a function (or distribution) $f(\theta,\phi)$ on the sphere based on \eqref{S2resun}   leads to 
the 2$\times$2 matrix operator
\begin{equation}
\label{qtfrhorS2}
f \mapsto A_f = \int_{\mathbb{S}^2} f(\theta,\phi) \rho_{r}(\theta,\phi) \,\frac{\sin\theta\,\ud \theta\, \ud \phi}{2\pi} = \begin{pmatrix}
  \lg f\rg + r\,C^{\mathbb{S}^2}_c (f)  &   r\,C^{\mathbb{S}^2}_s(f) \\
r\,\left(C^{\mathbb{S}^2}_s (f )\right)^{\ast}   &  \lg f \rg  - r\,C^{\mathbb{S}^2}_c(f)
\end{pmatrix}\,,
\end{equation}
where $ \lg f\rg:= \frac{1}{4\pi}\int_{\mathbb{S}^2}f(\theta,\phi)\,\sin\theta\,\ud \theta\, \ud \phi$ is the average of $f$ on the unit sphere and  $C^{\mathbb{S}^2}_c$ and $C^{\mathbb{S}^2}_s$ are  Fourier coefficients of $f$ on the sphere defined as
\begin{equation}
\label{CcCsS2}
C^{\mathbb{S}^2}_c(f) = \frac{1}{4\pi}\int_{\mathbb{S}^2}f(\theta,\phi)\,\cos\theta\,\sin\theta\,\ud \theta\, \ud \phi\, , \quad C^{\mathbb{S}^2}_s(f) = \frac{1}{4\pi}\int_{\mathbb{S}^2}f(\theta,\phi)\,e^{i\phi}\, \sin^2\theta\,\ud \theta\, \ud \phi\, . 
\end{equation}
Since the sphere is a phase space with canonical coordinates $q\equiv \phi$, $p\equiv \cos\theta$, and $\ud q\,\ud p=\sin\theta\,\ud \theta\, \ud \phi$,  the  latter may be thought as the simplest functions to be quantized. We find for the quantization of $q$
\begin{equation}
\label{qtfrhorS2}
 A_{q}= \pi \begin{pmatrix}
  1  &   - i\,\frac{r}{4}\\
i\,\frac{r}{4}  &  1 
\end{pmatrix} = \pi + \frac{\pi \, r}{4}\sigma_2\,.
\end{equation}
 Its eigenvalues are $\pi   \pm \dfrac{\pi r}{4}$ with corresponding  eigenvectors $\binom{1}{\pm i}$. 
Its lower symbol is given by the smooth function
\begin{equation}
\label{lowsqS2}
\check{q}(\theta,\phi) = \pi - \frac{\pi r^2}{4}\sin\theta\sin\phi\,. 
\end{equation}
The quantization of $p$ yields the diagonal matrix
\begin{equation}
\label{ptfrhorS2}
 A_{p}= \frac{r}{3} \begin{pmatrix}
  1  & 0 \\
0  & - 1 
\end{pmatrix} =  \frac{ r}{3}\sigma_3\,, 
\end{equation}
with immediate eigenvalues $\pm \dfrac{r}{3}$ and lower symbol
 \begin{equation}
\label{lowspS2}
\check{p}(\theta,\phi) = \frac{\pi r^2}{3}\cos\theta\,. 
\end{equation}
Finally we note the commutation rule 
\begin{equation}
\label{crqpS2}
\left\lbrack A_q, A_p\right\rbrack = i\, \frac{\pi r^2}{6} \sigma_1\, . 
\end{equation}

\section{The example of the plane}
\label{plane}
The measure set is the euclidean plane (or complex plane) equipped with its uniform (Lebesgue)  measure
\begin{equation}
\label{measS2}
X= \R^2 \sim \C\, , \quad \mathrm{d}\nu(x) = \frac{\ud^2 z}{\pi}= \frac{\,\ud q\,\ud p}{2\pi} \, , \quad \, z= \frac{q+ip}{\sqrt{2}} \in \C\, .
\end{equation}
The group $G$ is the  Weyl-Heisenberg group $G_{\mathrm{WH}} = \{(\vs,z)\, , \, \vs\in \R\, , \, z\in \C\}$ with multiplication law
\begin{equation}
\label{WHlaw}
(\varsigma,z)(\vs^{\prime},z^{\prime})= (\vs+\vs^{\prime} + \mathrm{Im}(z\bar z^{\prime}), z+z^{\prime})\,. 
\end{equation}
In this group context, the plane $\C$ is viewed as  the coset $X= G_{\mathrm{WH}}/C\sim \C$ where $C$ is the center in the group $C=\{ (\vs,0)\, , \, \vs \in \R\}$.
Let $\mathcal{H}$ be a separable (complex) Hilbert space  with orthonormal basis $e_0,e_1,\dots, e_n \equiv |e_n \rangle, \dots$.  
Let us suppose that the basis element $|e_n\rg$ is a state for $n$ excitations of an harmonic system, e.g. a Fock number state $|n\rg$ for the quantum electromagnetic field with single-mode photons and for which $X= \R^2$ is the  plane of quadratures.   
Given an elementary quantum energy, say $\hbar \omega$, and a temperature $T$ (e.g. a noise one, like in electronics), a Boltzmann-Planck $T$-dependent density operator, i.e. \textit{thermal state} \cite{helstrom76}, is introduced as:
 \begin{equation}
\label{plboltrho}
\rho_T= \left( 1- e^{-\tfrac{\hbar \omega }{k_B T}}\right)\sum_{n=0}^{\infty} e^{-\tfrac{n\hbar \omega }{k_B T}}|e_n \rg\lg e_n|\,. 
\end{equation} 
We notice that at zero temperature\footnote{On the other hand, at high temperature or equivalently in the classical limit $k_BT \gg \hbar \omega$, and from a classical probability point of view, one notices that  we have the Rice probability density function \cite{helstrom76}. This Rice distribution is also obtained in an analogous fashion in a classical optics context (classical but probabilistic) : ``a constant phasor plus a random phasor sum'' which one may take to be the classical version of the quantum ``oscillator with a coherent signal superimposed on thermal noise'' (see the classical probabilistic description in \cite{goodman00}}, this operator reduces to the projector on the first basis element (``ground state'' or ``vacuum''),
\begin{equation}
\label{zeroT}
\rho_0 = |e_0\rg\lg e_0|\, .
\end{equation}
Introducing lowering and raising operators $a$ and $a^\dag$,
\begin{equation}
\label{lowraisop}
  a\, |e_n\rg  = \sqrt{n} |e_{n-1}\rg\, , \quad  a|e_0\rg = 0 \, , \quad a^{\dag} \,|e_n\rg  = \sqrt{n +1} |e_{n+1}\rg\, ,
\end{equation}  
which  obeys the canonical commutation rule,
  \begin{equation}
\label{ccr}
  [a,\adg]= I\,,
\end{equation}
we obtain  the number operator: $N= \adg a$ whose the spectrum is $\mathbb{N}$, with corresponding eigenvectors the basis elements, $N|e_n\rg = n |e_n\rg$.  
Having in hand these two operators, we build a  unitary irreducible representation of the Weyl-Heisenberg group through  the map :
\begin{equation}
\label{dispop}
G_{\mathrm{WH}}/C\sim \C \ni z \mapsto D(z) = e^{z\adg -\bar z a}\,  , \quad D(-z) = (D(z))^{-1} = D(z)^{\dag}\, ,
\end{equation}
and the composition law
\begin{equation}
\label{dzdzp}
D(z)D(z') = e^{\frac{1}{2}(z\bar{z'} -\bar{z} z')}D(z+z') = e^{(z\bar{z'} -\bar{z} z')}D(z')D(z) \,,
\end{equation}
which show that the map $z\mapsto D(z)$ is a projective unitary representation of the abelian group $\C$. 
Then, one easily derives from the Schur lemma  or directly that the family of displaced operators 
\begin{align}
\label{rhoTz}
\nonumber \rho_T(z)&:= D(z)\rho_TD(z)^\dagger\\
&= (1-t)\sum_{m,m^{\prime}} \left\lbrack \sum_{n}t^n \, D_{nm}(z)\,D_{m^{\prime}n}(-z)\right\rbrack\,|e_m\rg\lg e_{m^{\prime}}|\, , \quad t= e^{-\tfrac{\hbar \omega }{k_B T}}\, ,
\end{align}
where  the matrix elements $D_{m n}(z) $ of the operator $D(z)$ are given in terms of associated Laguerre polynomials $L^{(\alpha)}_n(t)$ \cite{helstrom76},
\begin{equation} 
\label{matelD1}
\lg e_m|D(z)|e_n\rg:= D_{m n}(z)  =\left(D_{nm}(-z) \right)^{\ast} = \sqrt{\dfrac{n!}{m!}}\,e^{-\vert z\vert^{2}/2}\,z^{m-n} \, L_n^{(m-n)}(\vert z\vert^{2})\, ,   \quad \mbox{for} \ m\geq n\,,
 \end{equation}
with $L_n^{(m-n)}(t) = \frac{m!}{n!} (-t)^{n-m}L_m^{(n-m)}(t)$ for $n\geq m$. With these properties, \eqref{rhoTz} reads more explicitly as
\begin{equation}
\label{rhoTz1}
\nonumber \rho_T(z)=   \rho_T +  (1-t)\sum_{m \neq m^{\prime}} \left\lbrack \sum_{n}t^n \, D_{nm}(z)\,D_{m^{\prime}n}(-z)\right\rbrack\,|e_m\rg\lg e_{m^{\prime}}|\, .
\end{equation}
 Resolution of  identity follow from the results given in Section \ref{covquant},
\begin{equation}
\label{residrhoTz}
\int_{\mathbb{C}} \, \rho_T(z) \,\frac{\ud^2 z}{\pi}= I\, . 
\end{equation}
More general constructions and results are given in \cite{bergaz14}. At zero temperature, we recover the standard (Sch\"odinger, Klauder, Glauber, Sudarshan) coherent states 
\begin{equation}
\rho_0(z):= |z\rg\lg z|\, , \quad |z\rg = D(z)|e_0\rg \,.
\end{equation}
Let us evaluate the probability distribution $p_{z_0;T}(z)$ issued from $\rho_T(z)$. The expression of $p_{z_0;T}(z)$ is rather elaborate
\begin{align}
\label{pz0z}
\nonumber p_{z_0;T}(z) &= \mathrm{tr}\left(\rho_T(z_0)\rho_T(z)\right)= \left(1- t\right)^2\, e^{-\vert z-z_0\vert^2}\times\\
 &\times \left\lbrack\sum_n t^{2n}\, \left(L_n^{(0)}(\vert z-z_0\vert^{2})\right)^2 +  2  \sum_{n^{\prime}>n} t^{n+ n^{\prime}}\, \frac{n}{n^{\prime}}\, \vert z-z_0\vert^{2(n^{\prime}-n)}\, \left(L_n^{(n^{\prime}-n)}(\vert z-z_0\vert^{2})\right)^2\right\rbrack\, .
\end{align}
 The first term in the sum can be given a compact form \cite{magnus66}\footnote{Warning: there are errors in Poisson generating function for Laguerre polynomials, correct formula is found in \textcolor{hyptxt}{\href{http://en.wikipedia.org/wiki/Laguerre\_polynomials}{WikiLaguerre}}}:
\begin{equation}
\label{1termsum}
\sum_n t^{2n}\, \left(L_n^{(0)}(\vert z-z_0\vert^{2})\right)^2= \frac{e^{-\vert z-z_0\vert^2\frac{t^2}{1-t^2}}}{\left(1- t^2\right)}\, I_0\left(\frac{2t\vert z-z_0\vert^2}{1- t^2}\right)\, ,
\end{equation}
where $I_0$ is a modified Bessel function. 
At $z=z_0$ \eqref{pz0z} reduces to
\begin{equation}
\label{pz0z0}
p_{z_0;T}(z_0)= \mathrm{tr}\rho_T^2(z_0)= \frac{1- t}{1+t}\, . 
\end{equation}
As expected, at zero temperature this quantity is equal to 1. It vanishes at infinite temperature. The pseudo-distance  \eqref{psdist1}
takes the form
\begin{equation}
\label{distzz0}
\delta(z_0,z) = \vert z-z_0\vert + n_T(\vert z-z_0\vert)\, , 
\end{equation}
where the $T$-dependent $n_T$ goes to 0 as $T\to 0$. It is only in the limit CS case that this quantity acquires its true euclidean distance meaning.  As for $d_{\mathrm{HS}}$, we get
\begin{equation}
\label{dhsplane}
d_{T;\mathrm{HS}}(z,z^{\prime})=\sqrt{ \mathrm{tr}(\rho_T^2(z) - \rho_T^2(z^{\prime})) }= \sqrt{2}\sqrt{\left(\frac{1-t}{1+t}\right)^2- p_{z;T}(z^{\prime}) }\,. 
\end{equation}

The quantization map based on $\rho_T(z)$ is given by
\begin{equation}
\label{quantrhoT}
f\mapsto A_f = \int_{\mathbb{C}} \, \rho_T(z) \,  f(z) \, \frac{\ud^2 z}{\pi}\, . 
\end{equation}
There are translational and rotational covariances.  Covariance w.r.t. complex translations reads as
\begin{equation}
\label{covtrans}
A_{f(z-z_0)} = D(z_0) A_{f(z)} D(z_0)^\dagger\, .
\end{equation}
To show rotational covariance, we  define in preamble the unitary representation $\theta \mapsto U_{\mathbb{T}}(\theta)$ of the torus $\SN^1$ on the Hilbert space $\mathcal{H}$ as the diagonal operator  
\begin{equation}
\label{unrotrep}
U_{\mathbb{T}}(\theta)|e_n\rg = e^{i (n + \nu) \theta}|e_n\rg\, , 
\end{equation}
 where $\nu$ is arbitrary real. 
Then, from  the matrix elements  of $D(z)$ one proves easily the  rotational covariance property
\begin{equation}
\label{rotcovD}
U_{\mathbb{T}}(\theta)D(z)U_{\mathbb{T}}(\theta)^{\dag} = D\left(e^{i\theta}z\right)\, .  
\end{equation}
From the diagonal nature of $\rho_T$ we derive  the covariance of $A_f$ w.r.t. complex rotations in the plane,
\begin{equation}
\label{rotcovAf}
U_{\mathbb{T}}(\theta)A_f U_{\mathbb{T}}(-\theta)= A_{\varrho(\theta)f}
\end{equation}
where $\varrho(\theta)f(z):= f\left(e^{-i\theta} z\right)$.
In particular, for the parity operator defined by
\begin{equation}
\label{parop}
{\sf P} = \sum_{n=0}^{\infty}(-1)^n|e_n\rg\lg e_n|\, , 
\end{equation}
we have
\begin{equation}
\label{parcov}
A_{f(-z)} = {\sf P} A_{f(z)} {\sf P}, \forall \,f\, .
\end{equation}
A covariance also holds for the conjugaison operator:
\begin{equation}
\label{conjcov}
  A_{\overline{f(z)}}= A_{f(z)}^\dagger, \forall \,f\,. 
\end{equation}
Canonical Commutation Rule is a $T$-independent outcome of the above quantization,
\begin{equation}
\label{regquant1}
A_{z} = a\varpi\left(0\right)-\left.\partial_{\bar{z}}\varpi\right\vert _{z=0}= a -\left.\partial_{\bar{z}}\varpi\right\vert _{z=0}\, .
\end{equation}
 Equivalently, with $z= (q+ip)/\sqrt 2$, 
\begin{equation}
\label{AqAp}
A_{q}  =\frac{1}{\sqrt{2}}\left(a+a^{+}\right)\equiv Q\,, \quad
A_{p}  =\frac{1}{\sqrt{2}i}\left(a-a^{+}\equiv P\right)\,.
\end{equation}
 From this their  commutator is canonical: 
\begin{equation}
\label{comqp}
A_{q}A_{p}-A_{p}A_{q}=i\left[a,a^{+}\right]=i I\,.
\end{equation} 
We now turn our attention to the simple quadratic expressions.
\begin{equation}
\label{quadraq}
 A_{q^2}=Q^2-\frac{s}{2} \, , \quad A_{p^2}=P^2 - \frac{s}{2} \,,
\end{equation} 
where $s := - \coth\dfrac{\hbar \omega}{2k_B T}$.
It follows that 
\begin{equation}
\label{quantosc2}
A_{\vert z \vert^2} = \adg a + \frac{1-s}{2}\,,
\end{equation}
where $\vert z \vert^2 $ is the energy (in appropriate units) for the harmonic oscillator. 
The difference between the ground state energy
 $E_0= (1-s)/2$, and the minimum of the quantum potential energy  $E_m=[\min(A_{q^2}) + \min(A_{p^2})]/2 = - s/2$  is independent of the  temperature, namely $E_0-E_m=1/2$ (experimentally verified in 1925). 
It has been proven in \cite{bergayou13} (at least in the CS case) that these constant shifts in energy are inaccessible to measurement.

We now turn  our attention to the quantization of the  angle or phase.
We write $z = \sqrt J\, e^{i\gamma}$ in action-angle $(J,\gamma)$ notations for the harmonic oscillator. The quantization of a function $f(J, \gamma)$ of the action $J\in \R^+$ and  of the  angle $\gamma= \arg(z)\in [0,2\pi)$, which is $2\pi$-periodic in $\gamma$,  yields formally the operator
\begin{equation}
\label{aaquanta}
A_{f} = \int_0^{+\infty}\ud J \int_0^{2\pi}\frac{\ud\gamma}{2\pi} f(J,\gamma)\rho_T\left(\sqrt{J}e^{i\gamma}\right)\,. 
\end{equation}
The angular covariance property takes the form
\begin{equation}
\label{covquantaa}
U_{\mathbb{T}}(\theta)A_f U_{\mathbb{T}}(-\theta)= A_{T(\theta)f}\, , \quad T(\theta)f(J,\gamma) := f(J, \gamma -\theta)\, . 
\end{equation}
 In particular, let us quantize  the discontinuous $2\pi$-periodic angle function $\gimel(\gamma) = \gamma$ for $\gamma \in [0, 2\pi)$. Since this angle function is real and bounded,  its quantum counterpart $A_{\gimel}$  is a bounded self-adjoint operator, and it is covariant according \eqref{covquantaa}. In the basis $|e_n\rg$, it is given by  the infinite matrix:
\begin{equation}
\label{scsphaseop}    
A_{\gimel}= \pi\,  1_{{\mathcal H}} + i \, \sum_{m\neq m^{\prime}}{\sf F}_{mm^{\prime}}(t)\, \frac{1}{m^{\prime}-m}\, |e_m\rg\lg e_{m^{\prime}}|\, ,
\end{equation}
where
\begin{equation}
\label{Fmm'}
{\sf F}_{mm^{\prime}}(t)= (1-t)\frac{\Gamma\left(\frac{m+m^{\prime}}{2}+1\right)}{\sqrt{m m^{\prime}}}\, (1-t)^{\frac{m^{\prime}-m}{2}}\, {}_2F_1\left(-m,\frac{m^{\prime}-m}{2};- \frac{m+m^{\prime}}{2};t\right)
\end{equation}
is  symmetric w.r.t. permutation of $m$ and $m^{\prime}$ (from the well-known ${}_2F_1\left(a,b;c;x\right)= (1-x)^{c-a-b}
{}_2F_1\left(c-a,c-b;c;x\right)).$ 

This operator has spectral measure with support $[0,2\pi]$.  For a detailed study of such an operator in the CS case ($T=0=t$), see \cite{bergaz14} and in the case $T>0$ see \cite{gasza14}.

\section{The example of the half-plane}
\label{halfplane}
The measure set is the half plane  equipped with its uniform (Lebesgue)  measure
\begin{equation}
\label{measS2}
X= \R^+_{\ast}\times\R\equiv\Pi_{+}\, , \quad \mathrm{d}\nu(x) = \ud q\,\ud p \, , \quad \, q\in (0,+\infty)\, , \quad p\in \R\, . 
\end{equation}
Together with the multiplication $(q,p)(q_{0},p_{0})=(qq_{0},p_{0}/q+p),\, q\in\mathbb{R}_{+}^{\ast},\, p\in\mathbb{R}$,
$\Pi_{+}$ is viewed as the affine group Aff$_{+}(\mathbb{R})$ of
the real line. Aff$_{+}(\mathbb{R})$ has two non-equivalent UIR \cite{gelnai47,aslaklauder68}.
Both are square integrable and this is the rationale behind \textit{continuous
wavelet analysis} (see references in \cite{aagbook13}). The UIR $U_{+}\equiv U$
is realized in the Hilbert space $\mathcal{H}=L^{2}(\mathbb{R}_{+}^{\ast},\mathrm{d}x)$:
\begin{equation}
U(q,p)\psi(x)=(e^{ipx}/\sqrt{q})\psi(x/q)\,.\label{affrep+}
\end{equation}
In the same Boltzmann-Planck line as for the plane, we build the temperature-dependent density operator 
 \begin{equation}
\label{plboltrhohp}
\rho_T= \left( 1- t\right)\sum_{n=0}^{\infty} t^n|e_n \rg\lg e_n|\,, \quad t = e^{-\tfrac{\hbar \omega }{k_B T}}\, ,  
\end{equation}  
where $\{|e_n\rg\, |\, n\in \mathbb{N}\}$ is an orthonormal basis of $\mathcal{H}$. Let us choose that one which is  built from Laguerre polynomials,
\begin{equation}
\label{LagOB}
e_n \leftrightarrow e_n(x) = \sqrt{\frac{n!}{\Gamma(n+\alpha+1)}}\, e^{-\frac{x}{2}}\, x^{\frac{\alpha}{2}}\, L_n^{(\alpha)}(x)\, , \ \int_{0}^{\infty}e_n(x)\, e_{n^{\prime}}(x) \ud x = \delta_{n n^{\prime}}\,, 
\end{equation}
where $ \alpha > -1$ is a free parameter. Then, from \cite{magnus66}, the operator $\rho_T$ acts on $\mathcal{H}=L^{2}(\mathbb{R}_{+}^{\ast},\mathrm{d}x)$ as the integral transform
\begin{equation}
\label{intransf}
\rho_T: \psi(x) \mapsto \rho_T(\psi)(x) = \int_{0}^{\infty} \mathcal{K}_T(x,y)\,\psi(y)\, \ud y\, , 
\end{equation}
where the integral kernel is given by
\begin{equation}
\label{intkerLag}
\mathcal{K}_T(x,y) =(1-t) t^{-\alpha/2}\, e^{-\frac{1}{2}\frac{t}{1-t}(x+y)}\, I_{\alpha}\left(2\frac{\sqrt{t x y}}{1-t}\right)\, . 
\end{equation}
Again, one  derives from the Schur lemma  that the transported operators
\begin{equation}
\label{rhoTqp}
\rho_T(q,p):= U(q,p)\rho_TU(q,p)^\dagger
\end{equation}
 resolve the identity, 
\begin{equation}
\label{residrhoTqp}
\int_{\Pi_{+}}\rho_T(q,p) \,\frac{\mathrm{d}q\,\mathrm{d}p}{c_{\rho}}= I\, , 
\end{equation}
where the constant $c_{\rho}$ is obtained from the integral through standard calculations in wavelet theory,
\begin{align}
\label{croexpl}
\nonumber c_{\rho} &= \int_{\Pi_{+}}\lg e_0| \rho_T(q,p)|e_0\rg \,\mathrm{d}q\,\mathrm{d}p= (1-t) \int_{\Pi_{+}}\vert\lg e_0| U(q,p)|e_0\rg\vert^2 \,\mathrm{d}q\,\mathrm{d}p\\
&= 2\pi(1-t)  \int_{0}^{\infty}  (e_0(x))^2\, \frac{\ud x}{x}=(1-t) \frac{2\pi}{\alpha}\, . 
\end{align}
The resolution of the identity imposes the painless restriction $\alpha >0$ and reads finally
\begin{equation}
\label{residrhoTqpF}
\frac{\alpha }{(1-t)}\int_{\Pi_{+}}\rho_T(q,p) \,\frac{\mathrm{d}q\,\mathrm{d}p}{2\pi}= I\, .
\end{equation}
We leave the main results of the corresponding quantization to a future publication \cite{gatalal14}. 

\section{Conclusion}
\label{conclusion}

We would like to conclude with a few words about the relation between formalism based on POVM, regardless of whether it is used in a quantization context like here, or in  quantum measurement, and (quantum) statistical  inference.  Full developments will be  the subject for a separate paper.
For that, (Bayesian or not), as we see it, one needs to start from, (first and foremost), a context which includes a source of data which would be modeled by the use of a family of PV measures indexed by some parameters of theoretical interest (lying in space $(X,\nu)$) for the system being studied.
Then, given observed data, one turns it around and constructs a ( ``posterior'' or ``inferred'') probability model for the parameters in $X$, of theoretical interest, which is based upon a POV measure on $X$. The nice thing about coherent states, of course, and more generally density operators as they were built in the present paper,  is that they do the job in both directions.
 POV measure is generally conceived of as an attribute of quantum physics in contrast to classical physics. In other words,  POV measure is considered to be a generalization of PV measure (which of course it is mathematically) which one needs when generalizing from classical to quantum physics. But, to us, POV measure is an attribute of probabilistic inference (classical or quantum) and an associated PV measure models the accompanying source of data.
But note, if one has a deterministic model for the physics, none of the above applies. One has a direct route, provided by theory, from data to the parameters of interest. No need to make much of a distinction. Let us take a simple  example from Medecine\&Biology. Small amounts of dopamine obtained from brain tissue can be measured by preparing a fluorescent derivative. In order to connect the fluorescence measurement with the amount of dopamine one can run “standards”.  No problem. As long as there is a deterministic connection between the two.
The problem of inference comes up when we have probability modeling rather than deterministic. In that sense one can say it is quantum rather than classical; except, as we all know, there are classical contexts in which we need to use a probability model. In that case, POV measure would also apply in a classical situation.
Theoreticians, (it seems to us), are of course primarily interested in the parameters of theoretical interest for a particular system being studied so, in that case, one focuses upon the POV measures and doesn't necessarily include the other part (the data part).
As an aside: a famous American baseball player and homegrown philosopher once said: ``being a good (baseball) pitcher is 90\% mental and the other half is physical''.
Perhaps theoretical physics (``mental” POVM) bears a similar relationship to the (all important) physical. 

\appendix

%\section{Quantum world from classical probabilistic world: a minimal non trivial case}
%\label{N6n2}
\section{Parametrisations of $2\times2$ real density matrices}
\label{pararho}
%\subsection{}
There are various expressions for a density matrix acting on the Euclidean plane, i.e. a $2\times2$ real positive matrix with trace equal to 1. The most immediate one is the following with parameters $a$ and $b$: 
\begin{equation}
\label{dens-a-b}
\rho := {\sf M}(a,b) = \begin{pmatrix}
   a   &  b \\
    b  &  1-a
\end{pmatrix}\, , \quad 0\leq a\leq 1\, , \quad \Delta:= \det \rho = a(1-a)-b^2 \geq 0\,. 
\end{equation}
The above inequalities imply the following ones
 \begin{equation}
\label{ineqab}
 0\leq a(1-a)\leq \frac{1}{4}\, , \quad 0\leq \Delta \leq \frac{1}{4}\, , \quad   -\frac{1}{2} \leq b \leq \frac{1}{2}\,. 
\end{equation}
Let 
\begin{equation}
\label{eigrho}
\frac{1}{2} \leq \lambda = \frac{1}{2}(1+\sqrt{1-4\Delta}) \leq 1
\end{equation}
 be the highest eigenvalue of $\rho$ (the lowest one is $0\leq 1-\lambda\leq 1/2$). The spectral decomposition of $\rho$ reads as
\begin{equation}
\label{spedec}
\rho = \lambda |\phi\rg\lg\phi| + (1-\lambda) \left|\phi + \frac{\pi}{2}\right\rg\left\lg\phi+ \frac{\pi}{2}\right| \end{equation}
where 
\begin{equation}
\label{vectphi}
|\phi\rg \equiv \begin{pmatrix}
    \cos\phi      \\
    \sin\phi    
\end{pmatrix} \, , \quad - \frac{\pi}{2}\leq \phi\leq \frac{\pi}{2}\, ,
\end{equation}
is the corresponding unit eigenvector, chosen as pointing in the right half-plane. We could have as well chosen the opposite $|\phi + \pi\rg= -|\phi\rg$ pointing in the left half-plane since $|\phi + \pi \rg\lg\phi + \pi| = |\phi\rg\lg\phi|$. Our choice corresponds to most immediate in terms of orthonormal basis of the plane issued from the canonical one $\{ |0\rg\, , \, |\pi/2\rg\}$ through the rotation by $\phi$. 

Let us make explicit the decomposition \eqref{spedec}, 
\begin{equation}
\label{spedec1}
\rho = \begin{pmatrix}
  \left(\lambda-\frac{1}{2}\right)\cos(2\phi) + \frac{1}{2}    &    \left(\lambda-\frac{1}{2}\right)\sin(2\phi) \\
   \left(\lambda-\frac{1}{2}\right)\sin(2\phi)    &    \left(\frac{1}{2}- \lambda\right)\cos(2\phi) + \frac{1}{2} 
\end{pmatrix}\, . 
\end{equation}
We derive from this expression the polar parametrization of the $(a,b)$ parameters of $\rho$:
\begin{equation}
\label{polpar-a-b}
a-\frac{1}{2} =  \left(\lambda-\frac{1}{2}\right)\cos(2\phi) \, , \quad b = \left(\lambda-\frac{1}{2}\right)\sin(2\phi)\, . 
\end{equation}
In return, we have the angle $\phi\in [-\pi/2,\pi/2]$ in function of $a$ and $b$
\begin{equation}
\label{phiab}
\phi = \left\lbrace\begin{array}{cc}
 \frac{1}{2} \arctan\frac{b}{a-1/2} \, ,     &  - \frac{\pi}{4}\leq \phi\leq \frac{\pi}{4} \, ,  \\
       \frac{1}{2} \arctan\frac{b}{a-1/2}  + \frac{\pi}{4} \, ,   & \vert \phi\vert \geq \frac{\pi}{4}    \, . 
\end{array}   \right.
\end{equation}
In this way, each $\rho$ is univocally (but not biunivocally) determined by a point in the  unit disk, with polar coordinates $(r:=2\lambda-1, \Phi:=2\phi)$, $0\leq r\leq1$, $-\pi\leq \Phi <\pi$. 

Also note the alternative expression issued from \eqref{spedec1}:
\begin{equation}
\label{spedec2}
\rho \equiv {\sf R}(r,\Phi)=  \frac{1}{2}(I +   r \,\mathcal{R}(\Phi) \sigma_3)= \frac{1}{2}(I +     (2\lambda-1)\,\mathcal{R}(\phi) \sigma_3\mathcal{R}(-\phi))\, ,  
\end{equation}
where $\mathcal{R}(\Phi)$ is the rotation matrix in the plane
\begin{equation}
\label{rotmat}
\mathcal{R}(\Phi) = \begin{pmatrix}
\cos\Phi &  -\sin\Phi\\
   \sin\Phi    &    \cos\Phi
\end{pmatrix}\, ,
\end{equation}
and $\sigma_3$ is the diagonal Pauli matrix
\begin{equation}
\label{pauli3}
\sigma_3 = \begin{pmatrix}
1 & 0\\
0 &  -1
\end{pmatrix}\, .
\end{equation}
Note the important property used to get the second equality in \eqref{spedec2}:
\begin{equation}
\label{rotmatsig}
\mathcal{R}(\Phi) \sigma_3= \begin{pmatrix}
\cos\Phi &  \sin\Phi\\
   \sin\Phi    &  -  \cos\Phi
\end{pmatrix}= \sigma_3\mathcal{R}(-\Phi) \, . 
\end{equation}
Therefore the expression of a matrix density  to which we shall refer mostly often through the paper 
reads
\begin{equation}
\label{standrho}
\rho\equiv {\sf R}(r,\Phi) = \begin{pmatrix}
  \frac{1}{2}  + \frac{r}{2}\cos\Phi  &   \frac{r}{2}\sin\Phi  \\
\frac{r}{2}\sin\Phi     &   \frac{1}{2}  - \frac{r}{2}\cos\Phi
\end{pmatrix}\, . 
\end{equation}
From \eqref{spedec2} and \eqref{rotmatsig} we derive the interesting multiplication formula
\begin{equation}
\label{multrho}
\rho \rho'= {\sf R}(r,\Phi) {\sf R}(r',\Phi^{\prime}) = \frac{1}{2}\left( {\sf R}(r,\Phi) + {\sf R}(r',\Phi^{\prime}) +\frac{rr'}{2} \mathcal{R}(\Phi-\Phi^{\prime}) -\frac{I}{2}\right)\, , 
\end{equation}
and the resulting  (non-closed!) ``algebra'' of real density matrices, 
\begin{equation}
\label{algrho}
[\rho,\rho']= -irr'\sin(\Phi -\Phi^{\prime}) \sigma_2\, , \quad \{\rho,\rho'\} = \rho + \rho' + (\cos(\Phi - \Phi^{\prime})- 1/2)I\, . 
\end{equation}
\subsection{Covariance}

The expression \eqref{standrho} is convenient to examine the way a density matrix transforms under a rotation $\mathcal{R}(\omega)$ in the plane. We have, 
\begin{align}
\label{rotrho}
\nonumber \rho&\equiv {\sf R}(r,\Phi) \mapsto \mathcal{R}(\omega) {\sf R}(r,\Phi)\mathcal{R}(-\omega)=  \frac{1}{2}(I +     (2\lambda-1)\,\mathcal{R}(\phi + \omega) \sigma_3\mathcal{R}(-\phi - \omega))\\ &= {\sf R}(r,\Phi + 2\omega) \equiv \rho(\omega)\, . 
\end{align}

\subsection{Integrals of density matrix}
\label{integrho}
The computation of the three following (whose two are partial or marginal) integrals is straightforward:
\begin{equation}
\label{margtheta}
\frac{1}{\pi}\int_0^{2\pi} {\sf R}(r,\theta)\, \ud\theta= I\,.
\end{equation}
\begin{equation}
\label{margomega}
\frac{1}{\pi}\int_0^{2\pi} \rho(\omega) \, \ud\omega= \frac{1}{\pi}\int_0^{2\pi} {\sf R}(r,\theta + 2\omega)\, \ud\omega= I\,.
\end{equation}
\begin{equation}
\label{margr}
\int_0^{1}{\sf R}(r,\theta)\, r\ud r= \frac{1}{3}{\sf R}(1,\theta) + \frac{1}{12}I\,.
\end{equation}
\begin{equation}
\label{rtheta}
\frac{2}{\pi}\int_{\mathcal{D}}{\sf R}(r,\theta)\, \ud S= I\,,
\end{equation}
where $\mathcal{D}$ is the unit disk and $\ud S= r\ud r\ud\theta$. 

%\subsection{Resolution of the unity and quantization}
%\label{resunquant}
%\subsection{Quantization of functions on the unit circle}
%The resolution of the unity \eqref{margtheta} allows to quantize functions or distributions (if properly defined) on the unit circle along the linear map
%\begin{equation}
%\label{quantmapcirc}
%f(\theta) \mapsto A_f^{\rho} = \frac{1}{\pi}\int_0^{2\pi} f(\theta)\, {\sf R}(r,\theta)\, \ud\theta\, . 
%\end{equation}
%The matrix realization with respect to the basis $\{|0\rg,\, , |\pi/2\rg\}$ of  the operator $A_f$ reads as 
%\begin{equation}
%\label{matAfcirc}
%A_f^{\rho} = \begin{pmatrix}
% \langle f\rangle + r \langle f \cos(\cdot)\rangle   &  r \langle f \sin(\cdot)\rangle   \\
% r \langle f \sin(\cdot)\rangle    &   \langle f\rangle - r \langle f \cos(\cdot)\rangle
%\end{pmatrix}\, , \quad  \langle f\rangle:= \frac{1}{2\pi}\int_0^{2\pi} f(\theta)\, \ud\theta
%\end{equation} 
% 
%The most elementary example is the quantum angle.
% \begin{equation}
%A_{\theta}^{\rho}= \begin{pmatrix}
% \pi -b     &  \frac{1}{2}  - a\\
%   \frac{1}{2}  - a    &  \pi - b
%\end{pmatrix} \,, 
%\end{equation}
%with eigenvalues $\pi - b  \pm \left(\tfrac{1}{2}-a\right)$ and corresponding eigenvectors  $|\pi/4\rg$, $|3\pi/4\rg$.
%We exploit in the next section these elementary ideas and results.  
%%\subsection{Quantization of functions on the unit disk}

\section{SU(2) as unit quaternions acting in $\mathbb{R}^{3}$\label{sec:su2}}
\label{allSU2}
%Here, as a first example, we consider the case $G=$SU(2) and we pick
%for $U$ the spin one-half UIR. 

\subsection{Rotations and quaternions }

A convenient representation is possible thanks to quaternion calculus.
We recall that the quaternion field as a multiplicative group is $\mathbb{H}\simeq\mathbb{R}_{+}\times$SU(2).
The correspondence between the canonical basis of $\mathbb{H}\simeq\mathbb{R}^{4}$,
$(1\equiv e_{0},e_{1},e_{2},e_{3})$, and the Pauli matrices is $e_{a}\leftrightarrow(-1)^{a+1}i\sigma_{a}$,
with $a=1,2,3$. Hence, the $2\times2$ matrix representation of these
basis elements is the following: 
\[
\begin{pmatrix}1 & 0\\
0 & 1
\end{pmatrix}\leftrightarrow e_{0}\,,\,\begin{pmatrix}0 & i\\
i & 0
\end{pmatrix}\leftrightarrow e_{1}\equiv\hat{\imath}\,,\,\begin{pmatrix}0 & -1\\
1 & 0
\end{pmatrix}\leftrightarrow e_{2}\equiv\hat{\jmath}\,,\,\begin{pmatrix}i & 0\\
0 & -i
\end{pmatrix}\leftrightarrow e_{3}\equiv\hat{k}\,.
\]

Any quaternion decomposes as $q=(q_{0},\vec{q})$ (resp. $q^{a}e_{a},a=0,1,2,3$)
in scalar-vector notation (resp. in Euclidean metric notation). We
also recall that the multiplication law explicitly reads in scalar-vector
notation: $qq^{\prime}=(q_{0}q_{0}^{\prime}-\vec{q}\cdot\vec{q^{\prime}},q_{0}^{\prime}\vec{q}+q_{0}\vec{q^{\prime}}+\vec{q}\times\vec{q^{\prime}})$.
The (quaternionic) conjugate of $q=(q_{0},\vec{q})$ is $\bar{q}=(q_{0},-\vec{q})$,
the squared norm is $\Vert q\Vert^{2}=q\bar{q}$, and the inverse
of a nonzero quaternion is $q^{-1}=\bar{q}/\Vert q\Vert^{2}$. Unit
quaternions, i.e., quaternions with norm $1$, the multiplicative
subgroup isomorphic to SU(2), constitute the three-sphere $S^{3}$.

On the other hand, any proper rotation in space is determined by a
unit vector $\hat{n}$ defining the rotation axis and a rotation angle
$0\leq\omega<2\pi$ about the axis.\vskip 0.4cm {\centering \setlength{\unitlength}{1.2mm}
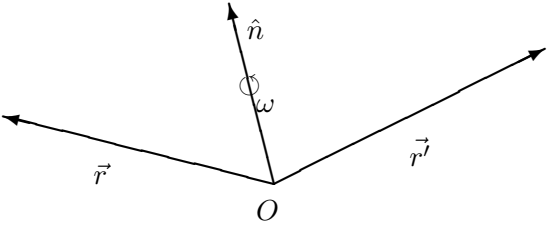
\begin{picture}(50,40) %\put(30,20){\vector(1,0){30}} 
%\put(30,20){\vector(4,1){20}} 
%\put(30,20){\vector(3,1){25}} 
\thicklines \put(30,10){\vector(2,1){30}} \put(28,6){$O$}
\put(45,12){$\vec{r'}$} %\put(30,20){\vector(1,2){10}} 
%\thicklines 
\put(30,10){\vector(-4,1){30}} \put(26,20){$\circlearrowleft$}
\put(28,18){$\omega$} \put(27,26){$\hat{n}$} \put(30,10){\vector(-1,4){5}}
\put(10,10){$\vec{r}$} %\thinlines 
%\put(30,20){\vector(-1,-1){5}} 
%\put(30,20){\vector(-1,-4){5}} 
\end{picture}} \vskip 0.4cm The action of such a rotation, $\mathcal{R}(\omega,\hat{n})$,
on a vector $\vec{r}$ is given by: 
\begin{equation}
\vec{r^{\prime}}\overset{def}{=}\mathcal{R}(\omega,\hat{n})\cdot\vec{r}=\vec{r}\cdot\hat{n}\,\hat{n}+\cos\omega\,\hat{n}\times(\vec{r}\times\hat{n})+\sin\omega\,(\hat{n}\times\vec{r})\,.\label{rotnot1}
\end{equation}

The latter is expressed in scalar-vector quaternionic form as 
\[
(0,\vec{r^{\prime}})=\xi(0,\vec{r})\bar{\xi}\,,
\]
where 
\[
\xi:=\left(\cos\frac{\omega}{2},\sin\frac{\omega}{2}\,\hat{n}\right)\in\mathrm{SU}(2)\,,
\]
or, in matrix form, 
\begin{align}
\xi & =\left(\begin{array}{cc}
\xi_{0}+i\xi_{3} & -\xi_{2}+i\xi_{1}\\
\xi_{2}+i\xi_{1} & \xi_{0}-i\xi_{3}
\end{array}\right)\nonumber \\
 & =\left(\begin{array}{cc}
\cos\frac{\omega}{2}+in^{3}\sin\frac{\omega}{2} & \left(-n^{2}+in^{1}\right)\sin\frac{\omega}{2}\\
\left(n^{2}+in^{1}\right)\sin\frac{\omega}{2} & \cos\frac{\omega}{2}-in^{3}\sin\frac{\omega}{2}
\end{array}\right)\,,\label{eq:xi-matrix}
\end{align}
in which case quaternionic conjugation corresponds to the transposed
conjugate of the corresponding matrix.

In particular, for a given unit vector 
\begin{align*}
\hat{n} & =(\sin\theta\cos\phi,\sin\theta\sin\phi,\cos\theta)\overset{def}{=}(\theta,\phi)\,,\\
 & 0\leq\theta\leq\pi\,,\quad0\leq\phi<2\pi\,,
\end{align*}
one considers the specific rotation $\mathcal{R}_{\hat{n}}$ that
maps the unit vector pointing to the north pole, $\hat{k}=(0,0,1)$,
to $\hat{n}$, 
\begin{equation}
\left(0,\hat{n}\right)=\left(0,\mathcal{R}(\theta_{\hat{n}},\hat{u}_{\phi_{\hat{n}}})\hat{k}\right)\equiv\xi_{\hat{n}}\left(0,\hat{k}\right)\bar{\xi}_{\hat{n}}\,,\quad\hat{u}_{\phi_{\hat{n}}}\overset{def}{=}(-\sin\phi_{\hat{n}},\cos\phi_{\hat{n}},0)\,,\label{rotspec}
\end{equation}
with 
\begin{equation}
\xi_{\hat{n}}=\left(\cos\frac{\theta_{\hat{n}}}{2},\sin\frac{\theta_{\hat{n}}}{2}\,\hat{u}_{\phi_{\hat{n}}}\right)\,.\label{xir}
\end{equation}

\vskip 0.4cm {\centering \setlength{\unitlength}{1.2mm} 
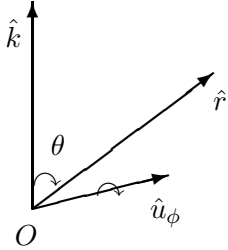
\begin{picture}(50,30)
\thicklines \put(30,10){\vector(4,3){20}} \put(30,10){\vector(4,1){15}}
\put(28,6){$O$} \put(50,21){$\hat{r}$} \put(43,9){$\hat{u}_{\phi}$}
\put(30,12.5){$\curvearrowright$} \put(37,11.5){$\curvearrowright$}
\put(32,16){$\theta$} \put(27,28){$\hat{k}$} \put(30,10){\vector(0,4){23}}
\end{picture}}

\subsection*{Acknowledgements}

Jean Pierre Gazeau thanks the CNPq for financial support, the  TWAS-ICTP (Trieste) and the CBPF (Rio)  for hospitality and support.

\end{document}